# Human breath analysis via cavity-enhanced optical frequency comb spectroscopy

Michael J. Thorpe, David Balslev-Clausen, Matthew S. Kirchner, and Jun Ye

JILA, National Institute of Standards and Technology and University of Colorado,

Boulder, CO 80309-0440, USA

Department of Physics, University of Colorado, Boulder, CO 80309-0390, USA

**To date, researchers have identified over 1000 different compounds contained in human breath. These molecules have both endogenous and exogenous origins and provide information about physiological processes occurring in the body as well as environment-related ingestion or absorption of contaminants[1,2]. While the presence and concentration of many of these molecules are poorly understood, many 'biomarker' molecules have been correlated to specific diseases and metabolic processes. Such correlations can result in non-invasive methods of health screening for a wide variety of medical conditions. In this article we present human breath analysis using an optical-frequency-comb-based trace detection system with excellent performance in all criteria: detection sensitivity, ability to identify and distinguish a large number of biomarkers, and measurement time. We demonstrate a minimum detectable absorption of $8 \times 10^{-10}$ cm$^{-1}$, a spectral resolution of 800 MHz, and 200 nm of spectral coverage from 1.5 to 1.7 μm where strong and unique molecular fingerprints exist for many biomarkers. We present a series of breath measurements including stable isotope ratios of $CO_2$, breath concentrations of CO, and the presence of trace concentrations of $NH_3$ in high concentrations of $H_2O$.**

Several methods of trace molecular detection have been applied to the problem of breath analysis, including optical detection[3,4], mass spectrometry[5,6], and electronic noses[6,7]. To understand the choice of optical detection as the preferred technique for breath analysis, we first evaluate the available techniques in the context of the system criteria. For instance, mass spectrometry (MS) is extremely sensitive and thus capable of detecting very small quantities of the analyte molecule. However, when several molecules are present, MS has difficulty identifying a single component of the mixture. To remedy this problem, MS is often used in conjunction with gas chromatography to separate out various components of a mixture prior to measurement. While these hybrid systems are highly sensitive and accurate for a large number of biomarkers, they are also large, complex, and require a long period of time to perform a measurement[8]. Conversely, electronic nose devices are typically inexpensive and perform measurements rapidly. However, these devices, which are designed to measure volatile organic compounds, have difficulty distinguishing and accurately measuring concentrations of individual molecules from the group they are designed to detect[9]. Optical detection provides a good compromise as a general approach that can be applied to many molecules. The unique absorption spectrum of each molecule allows accurate identification and concentration measurements of a single molecule in the presence of many others. Furthermore, cavity enhancement techniques permit highly sensitive detection in a matter of seconds.

As laser technology becomes less expensive, more compact, and more reliable, the number of optical detection systems designed to detect biomarkers has increased. The advent of tunable laser diodes, for instance, has led to systems operating in the near infrared (NIR) that measure the ratio of stable isotopes of carbon to detect *Helicobacter pylori*, a leading cause of peptic ulcers[3]. Another NIR system detects methylamine to study its correlation with liver and renal diseases[10]. Quantum cascade lasers, operating in the mid-infrared, have been used to detect ammonia for diagnosis of renal failure[11].

These lasers are also used to detect nitric oxide for diagnosing asthma[4]. Finally, both quantum cascade and lead salt lasers have been used to detect ethane which is produced in lipid peroxidation and by some forms of cancer[12,13,14]. While all of these systems provide robust and highly sensitive detections of their analyte molecules, many operate in relatively narrow spectral regions. Others require long times to perform sensitive detections over a large spectral region. Therefore, the number of molecules that can be studied or detected by a single system is rather limited.

Recent approaches to trace detection that use broad bandwidth optical frequency combs have attempted to address this problem by creating parallel detection schemes that in a single shot record large spectral bandwidths[15,16,17,18,19]. Frequency-comb-based systems also take advantage of the high peak intensities of their pulsed output to easily access any spectral region from UV to far infrared via nonlinear conversion[20,21,22]. Furthermore, the development of mode-locked femtosecond fiber lasers has led to robust frequency comb sources capable of continuous operations without user intervention[23]. In this work we couple the broad spectrum of a mode-locked fiber laser to an optical enhancement cavity to greatly enhance the detection sensitivity of breath samples. High spectral resolution is achieved within the entire 1.5 - 1.7 μm spectral region with the use of a virtually imaged phased array (VIPA) detector [24,18]. The relevant spectrum covers many biomarkers, such as ethane 1.68-1.7 μm, acetone 1.67-1.68 μm, methane 1.63-1.69 μm, ethylene 1.62-1.66 μm, carbon dioxide 1.54-1.64 μm, carbon monoxide 1.56-1.6 μm, ammonia 1.5-1.54 μm, methylamine 1.5-1.53 μm.

The cavity-enhanced optical frequency comb spectrometer for breath measurements is shown in Fig. 1. The spectrometer consists of two subsystems; a continuous-flow gas system delivers breath samples, calibration gases, and the reference gas ($N_2$) to the detection chamber, and an optical subsystem records absorption features that are used to determine molecular concentrations. In the scheme of comb-based cavity-enhanced absorption spectroscopy[15,16,17,19], the intra-cavity absorption signal is enhanced by the cavity multi-pass effect and is recovered by comparing the light transmitted from a high finesse optical cavity with and without absorption present. Every frequency comb component is coupled to a corresponding cavity resonance mode. For a single tooth of the frequency comb, the electric field $E_t$ transmitted from the cavity can be written as the sum[25],

$$\frac{E_t}{E_{inc}} = \sum_{n=1}^{\infty} (R-1) R^{n-1} e^{-(2n-1)\alpha L/2} e^{i\phi_n(t)}. \tag{1}$$

Here, L is the cavity length, R is the mirror reflectivity, n refers to the number of round trips a photon travels inside the cavity, $\alpha$ is the intracavity absorption, and $E_{inc}$ is the incident electric field. The phase term $\phi_n(t)$ determines the interference of the fields inside the cavity and is written as

$$\phi_n(t) = 2\pi(2n-1)\left[ L\frac{\nu(t)}{c} - (n-1)\beta\frac{L^2}{c^2} \right] . \tag{2}$$

Here $\nu(t)$ is the time-dependent frequency of one of the incident comb teeth and $\beta$ is the rate at which the incident frequency is swept. The intensity of the beam transmitted from the cavity is given by $I(t) = E_t(t)E_t^*(t)$. The cavity transmission measurements presented here are made with a slow detector that does not resolve the time dependence of I(t). Instead, the integrated cavity transmission is measured. The mirror reflectivity and the cavity length together determine the absorption sensitivity of the instrument and provide a relationship between the change in the transmitted light and the level of intracavity absorption.

**Spectrometer characterization**

A mode-locked $Er^{+3}$ fiber laser provides 100 nm spectrum from 1.5 to 1.6 μm with 40 mW of average power. For spectral measurements between 1.5 and 1.6 μm, the output of the laser is coupled directly into the optical enhancement cavity. For spectral measurements from 1.6 to 1.7 μm, we exploit the high peak intensity of the mode-locked laser output by using a Raman shifting amplifier to shift the spectrum and provide 300 mW of power (Fig. 2a). A simple flipper mirror is used for rapid switching between the two laser configurations. The laser is mode-matched into the cavity constructed from two mirrors with a 2 m radius of curvature and a peak reflectivity of R = 0.99989 at 1.6 μm. The repetition frequency of the laser $f_r$ is 100 MHz, matching the free spectral range (FSR) of the cavity. The cavity mirrors are glued onto the ends of a pyrex tube forming an air-tight detection chamber. The mirror coatings are highly reflective and low dispersion, enabling sensitive detection over a 200 nm range from 1.5 to 1.7 μm (Fig. 2b).

Light transmitted from the cavity is dispersed in two dimensions by a VIPA spectrometer, consisting of a solid etalon oriented orthogonally to a diffraction grating[18,23]. The etalon angularly disperses frequency comb components in the vertical direction, creating a single vertical stripe of light. For wavelengths separated by greater than one free spectral range of the etalon, the grating disperses them horizontally, converting a single vertical stripe to many parallel stripes (or fringes). The dispersed light is then imaged onto an InGaAs photodiode array, each pixel containing the frequency-resolved intensity of the cavity transmission via different comb channels. Each frame recorded by the camera captures 25 nm of spectrum with a resolution of 800 MHz, corresponding to 3500 independent channels captured per frame. A 25 nm range of the spectrometer output is imaged onto the camera at a time, and a simple rotation of the grating provides access to other spectral regions. Eight separate measurements are acquired to cover the entire 200 nm spectral bandwidth provided by the laser/cavity. Figure 2c records eight independent measurements stitched together to provide an absorption spectrum of the air in our laboratory.

The minimum detectable absorption for the spectrometer is determined by comparing consecutive sets of images taken while the detection cavity contains only nitrogen. The differential intensity noise from these sets of images is then converted to absorption noise at a particular wavelength using the mirror reflectivity (Fig. 2b) and the integral of equation (1). Figure 3 shows the residual absorption noise at 1.6 μm as a function of the number of camera frames averaged. The contribution from the camera dark counts is also shown. The residual absorption noise is primarily due to laser intensity noise, cavity-laser coupling noise, and residual interference fringes generated within thin optical elements, such as the glass cover of the InGaAs diode array. The results show that after 2000 averages taken within 30 s, the minimum detectable absorption is $8 \times 10^{-10}$ $cm^{-1}$ per detection channel.

**Breath measurements**

The relevant molecules studied for breath analysis are CO, $CO_2$, $H_2O$, $CH_4$, and $NH_3$, with more than 8000 known spectral features between 1.5 μm and 1.7 μm. The large spectral complexity and congestion make it particularly challenging for optical trace detection to accurately identify specific target molecules from the forest of absorption features of other species. This challenge is intensified by the typical requirement that tiny absorptions arising from trace amounts of an analyte molecule must be found amongst absorptions of much larger concentrations of other molecules (such as $H_2O$ and $CO_2$) in the breath. The difference in the concentration of an analyte molecule and $H_2O$ or $CO_2$ can vary between $10^3$ and $10^9$. Nonetheless, the cavity-enhanced frequency-comb spectroscopy permits recognition of a great diversity of molecular fingerprints over a vast optical spectral region, resulting in accurate concentration measurements of multiple analytes within a breath sample.

Our protocol for breath measurements of $CO_2$ stable isotopes and CO concentration requires the test subject to first take one deep breath. Next, the subject inhales normally and the breath is held for five seconds. Finally, the first half of the breath is released into the air before the second half of the breath is exhaled into the 1 liter Tedlar sample bag shown in Fig. 1. The use of only the second half of the breath sample increases the concentration of alveolar breath that is measured.

The first set of our breath measurements involves the analysis of stable isotopes. Here we focus on measurements of $CO_2$ isotopes, but other isotopes such as $H_2O$ and $CH_4$ also exist in measurable quantities within our spectral region. Two spectral windows centred at 1.59 μm and 1.63 μm contain spectroscopic features of three isotopes of $CO_2$, all of roughly equal absorption strengths. Figure 4 shows the breath spectrum of a healthy graduate student in the window centred at 1.63 μm and illustrates a unique feature of broadband detection. The two zoomed-in panels (a) and (b) both contain a number of spectral features for each of the three isotopes. By measuring many absorption lines and computing a concentration for each line, the overall accuracy of the isotope ratio measurement is enhanced by the square root of the number of lines measured. To calculate the isotope ratios from the spectrum shown in Fig. 4c, a modest selection of 5 lines is chosen for each isotope. Typically, isotope measurements are expressed in parts per thousand such that $\delta = (R_{sample} - R_{standard})/R_{standard} \times 1000$, where $R_{standard}$ is the natural abundance of the isotope and $R_{sample}$ is the abundance found in the breath sample. Using this analysis we find that $\delta^{18}O = -9.1 \pm 4.2$ and $\delta^{13}C = -28.8 \pm 4.1$. The $\delta^{13}C$ ratio is currently the only breath biomarker that is used widespread for clinical diagnosis. However, the ability to measure other stable isotopes of $CO_2$ at increased accuracy, and the capability of measuring isotopes of other molecules such as $H_2O$ (1.55 μm) or $CH_4$ (1.65 μm), could significantly broaden the scope of future breath research.

Although CO has more than 170 spectral features between 1.55 and 1.61 μm, we found only one absorption line at 1.56474 μm that doesn't overlap with interfering absorptions from $H_2O$ and $CO_2$. The second breath measurement shown in Fig. 5a utilizes this spectral feature to detect the concentration of CO contained in the breath of two students. Student 1 is a smoker who had a cigarette 15 minutes prior to the beginning of the measurements while student 2 is a non-smoker. Both students provided five breath samples, each separated by 10 minutes. Both students exhibited constant levels of CO in their breath above the minimum detectable concentration of CO at 1.564 $\mu$m, which is 900 parts per billion (ppb). The average CO concentration measured for the smoker (6.5 ppm) was five times that of the non-smoker (1.3 ppm). While analyzing the data for the CO concentration, we noticed a small difference in the concentration of $CO_2$ exhaled by the two students (Fig. 5a). This observation led us to examine the sensitivity of our concentration measurements to possible variations in our measurement protocol. In particular, we investigated both the CO and $CO_2$ breath concentrations as a function of time the test subjects held their breath prior to exhaling into the sample bag. Figure 5b shows these dynamic results. For $CO_2$, two separate measurements were taken at each hold time and five spectral features were used from each measurement for a total of ten samples to determine the concentration at each hold time. For CO, only a single measurement was taken at each hold time and consequently the error bars shown for CO correspond to the minimum detectable concentration of 900 ppb. This test shows that while the $CO_2$ concentration is quite sensitive to how long the test subject holds their breath, the CO concentration is not to within the statistical uncertainty of the measurement.

The final breath-related measurement we performed was to determine the minimum detectable concentration of $NH_3$, a biomarker for renal failure that can be detected within a large concentration of $H_2O$. The absorption spectrum of $NH_3$ overlaps heavily with $H_2O$ in the spectral region from 1.5 $\mu$m to 1.55 $\mu$m. Two spectral windows, one at 1.513 $\mu$m and the other at 1.517 $\mu$m were found to contain at least two strong $NH_3$ absorption lines that have minimal overlap with $H_2O$ absorptions (Fig. 6). To find these spectral windows and determine the minimum detectable $NH_3$ concentration, we measured the absorption spectrum of a calibrated sample of 4.4 ppm $NH_3$ in balanced nitrogen and compared this spectrum to both the lab air spectrum, and spectra generated from breath samples. The feature with the greatest $NH_3/H_2O$ absorption ratio was centred at 1.512231 μm. The minimum detectable concentration of $NH_3$ is considered to have a corresponding absorption feature that is twice the standard deviation of the measurement noise. According to this criterion, the minimum detectable concentration of $NH_3$ in the lab air at 25 % relative humidity was 18 ppb. Since our breath measurements nominally contain 3 to 4 times higher $H_2O$ concentration, small overlaps with $H_2O$ absorptions lead to slightly higher minimum detectable concentrations of ammonia in breath. Although this limit is still higher than detectable $NH_3$ present in the breath of most healthy individuals, it is sufficient to detect increased $NH_3$ present in the breath of people at early stages of renal failure (Fig. 6). Furthermore, if techniques are implemented to remove water vapor from the breath sample prior to measurement, such as flowing the breath sample through a cold (0 °C) glass tube, a much lower detection limit for $NH_3$ can be achieved.

**Discussion**

The measurements presented in the previous section provide only a modest glimpse of the capabilities of our cavity-enhanced optical-frequency-comb spectrometer. The large number of biomarkers that can be measured by the system coupled with sub-minute measurement times enable a cost effective way to accelerate research in breath analysis and to provide reliable clinical instrumentations. Fortunately, frequency comb technology is progressing very rapidly with ever improved spectral coverage, power output, and ease of use. Commercial mode-locked fiber lasers are available today with all operations performed by the laser completely automated. Furthermore, since such lasers are based on technology created for the telecommunications industry, they are reliable for years of uninterrupted operations.

Similarly, the detection cavity is highly resistant to contamination, demonstrating no degradation in finesse after hundreds of hours of continuous flow of gas and breath samples.

Frequency comb technology is currently being extended for convenient operation in the 3 μm spectral region that promises to increase the detection sensitivity for biomarkers by a factor of 10 to $10^3$ compared to our current system. Also, new designs of high-finesse prism cavities promise to expand the spectral coverage of optical cavities from 200 nm to several microns, greatly increasing the number of biomarkers that can be measured by a single system [26]. Although future developments of frequency comb-based systems promise to make further dramatic advances, we note that our current system can already be used for large scale clinical trials to gather statistics necessary to institute new, non-invasive, health screening tests.

**Methods**

**Gas handling system.** The gas handling system is operated under a condition of continuous flow created using a low pressure regulator and a scroll pump with a choke valve. Continuous flow of gas is used to minimize the effect of adsorption while maximizing the repeatability of absorption measurements and the throughput of the measured gas. The regulator provides a constant pressure inside the cavity while the scroll pump and choke valve allow control over the gas flow rate. In this configuration, flows of .1 to 10 liters/minute at pressures ranging from 100 to 400 torr are achievable. Both the Tedlar sample bag and the detection chamber have a volume of 1 liter and the typical flow rate of sample gas through the system was 2 liters/minute. The tubing, sample bag, and valves are made of teflon to minimize adsorption and to resist corrosion. The detection chamber is constructed from a pyrex tube with inlet and outlet ports for gas flow and is heated to 40 °C to protect against condensation and to minimize temperature dependent fluctuations of the cavity length.

**Noise mitigation.** To reduce the noise present in absorption measurements, the laser is robustly mode-locked with a stable frequency comb spectrum. The gas flow within the detection chamber is maintained at a sufficiently low rate (<10 liters/minute) such that the flow is laminar and intracavity pressure fluctuations are minimized. The detection cavity is held on vibration-isolating v-blocks and is enclosed in an acrylic box lined with sound-damping foam. To further minimize the effects of mechanical noise when coupling the comb modes to the cavity, the comb frequencies are dithered across the cavity modes at a rate of 1.5 kHz with an amplitude such that the comb is on resonance with the cavity for 1/3 of the total sweep time. The periodic build-up and ring-down signal is used to generate a locking error signal. The swept frequency comb is then stabilized to the cavity such that the comb modes come on resonance with the cavity at the same point in each sweep cycle. Locking in this manner acts to stabilize the fringe pattern generated by the VIPA and provide stable operations for many hours at a time.

**Spectrometer calibration.** The equation $\alpha(v) = n\, S_i\, g(v)$ relates the absorption coefficient measured by our spectrometer to the concentration of the gas under measurement. Here, $\alpha(v)$ is the absorption coefficient, $n$ is the gas concentration, $S_i$ is the line intensity, and $g(v)$ is the lineshape. The values for $S_i$ can either be measured or found from a variety of references[27,28,29]. However, the lineshape depends on actual experimental conditions. In particular, choosing the appropriate intra-cavity gas pressure is important for maximizing the molecular absorption without compromising the spectral resolution by collision broadening. To quickly determine the concentration of an analyte molecule, we have created a database of the centre frequencies, line intensities, and pressure broadening coefficients for many of the absorption features under study. For measurements presented in this article, an intracavity gas pressure of 250 torr is chosen such that the average linewidth for a gas under measurement is 1.8 GHz, resolved by the 800 MHz resolution of our system. A modified Voigt fit that includes the resolution of the spectrometer is performed to find the $g(v)$ and relate $\alpha(v)$ to *n*. By finding the value of the instrument resolution that provides the best fit to the experimentally recorded lineshape, an instrument resolution of 800 MHz has been experimentally verified.

To calibrate the spectrometer we use three gas mixtures containing a total of four trace gases. These mixtures consist of 10 ppm $CO_2$, 1 ppm $CH_4$, 1 ppm $C_2H_4$, and 4.4 ppm $NH_3$, all mixed with balanced $N_2$. The accuracy of the trace gas concentrations is specified by the manufacturer to within 1% of the quoted concentration. Since the spectral lines from these four molecules cover the entire spectral range from 1.5 μm to 1.7 μm, we are able to verify the calibration of the absorption loss provided by equation (1) over this entire spectral region.

**VIPA image analysis.** Recovering spectral absorption information from a pair of reference and absorption images requires several steps of computer processing. Frames are passed from the camera to the control computer at a 60 Hz rate. These images are acquired and averaged in real-time. The user presets the number of images to be averaged for both the reference and absorption pictures before beginning the image acquisition. Once the two sets of images have been acquired and averaged into a single pair of images, the spectral recovery process is initiated.

A flow diagram of the spectral recovery process is presented in Fig. 7. A pair of 256 x 320 pixel images, one reference and the other contains absorption, is converted into a 25 nm absorption spectrum. The first step in this process is to locate the nearly vertical fringes produced by the VIPA spectrometer within the picture frame. This is typically done using the reference image. Fringes are identified by an algorithm that starts at the peak of each fringe at the middle of the frame (i.e. row 128) and 'walks' along each fringe upward and downward, locating the peak of each fringe on the frame. Next, the intensity of each fringe is converted into a column vector, creating a fringe array containing all of the frequency resolved cavity transmission information.

As a first-order correction for laser intensity drift between acquisitions of the absorption and reference frames, an area of low absorption in the absorption fringe array is compared to the reference fringe array, and the absorption fringe intensity is equalized to the reference fringe intensity. The absorption fringe array is then subtracted from the reference fringe array, before being divided by the reference fringe array. The result is a differentially measured and normalized fringe array that contains the relative intensity information.

The etalon FSR is determined using the pattern of absorption peaks from the subtracted fringe array. All unique information about molecular absorptions is contained within this interval. The etalon used in our system has a FSR = 50 GHz. Once this interval is determined, the subtracted fringe array is unwrapped into a one dimensional spectrum of relative intensity versus wavelength. The generalized grating equation for the VIPA can be used to obtain a high precision mapping of position on the subtracted fringe array to wavelength[23]. A low pass filter is then employed to remove noise from the picture at spatial frequencies higher than the resolution of the spectrometer. Also, for sharply peaked spectra, a background subtraction routine is used to remove higher order intensity fluctuations between the reference and absorption frames due to laser noise, laser/cavity coupling noise, or pointing instability. Finally, the relative intensity information is converted into quantitative absorption strengths using the integral of equation (1).

**Acknowledgements** We acknowledge funding support from AFOSR, Agilent, DARPA, NIST, NSF, and Univ. of Colorado POCg. The authors thank S. Diddams, T. Schibli, D. Hudson, and K. Moll for useful discussions, and S. Rittenhouse and M. Stowe for donations of their breath samples.

**Figure Captions**

Fig. 1  Schematic of the cavity-enhanced direct-frequency-comb spectrometer, along with the gas handling system for breath analysis.

Fig. 2  a) The spectrum generated by the mode-locked fiber laser from 1.5 to 1.6 µm (red) and the amplified, Raman shifted, spectrum from 1.6 to 1.7 µm (blue).  b) The measured reflectivity of the cavity mirrors as a function of wavelength. c) The measured absorption spectrum of the air in our laboratory from 1.5 µm to 1.7 µm.

Fig. 3  The absorption detection noise floor as a function of the number of pictures averaged. The camera dark count noise is also shown.

Fig. 4  A breath spectrum between 1.622 µm and 1.638 µm.  Several windows in this region contain spectroscopic features of three isotopes of $CO_2$, with nearly equal absorption strengths. a) and b) Two zoomed-in spectral windows where line positions and intensities of relevant $CO_2$ transitions are shown. Besides $CO_2$ peaks, strong absorption features of $H_2O$ and $CH_4$ are detected in b).   c) The entire range over which this condition exists.

Fig. 5  a) The $CO_2$ and CO absorption spectra of student 1 (smoker) and student 2 (non-smoker) in the 1.564 µm spectral region, along with line intensities from the Hitran spectral database. The smoker's obvious increase in CO concentration is clearly detected. b) The breath concentration of $CO_2$ in parts per thousand (ppt) and CO in parts per million (ppm) as a function of time during which the test subject holds their breath prior to exhaling into the sample bag.

Fig. 6  The absorption spectrum of lab air superimposed over the spectrum of 4.4 ppm of $NH_3$.  Hitran line intensities for $H_2O$ are included to identify the absorption features

of the lab air spectrum.  The dashed grey line represents the typical concentration of NH$_3$ contained in the breath of a patient at final stages of renal failure.

Fig. 7  The software flow chart for generating a spectrum from a reference and an absorption image.

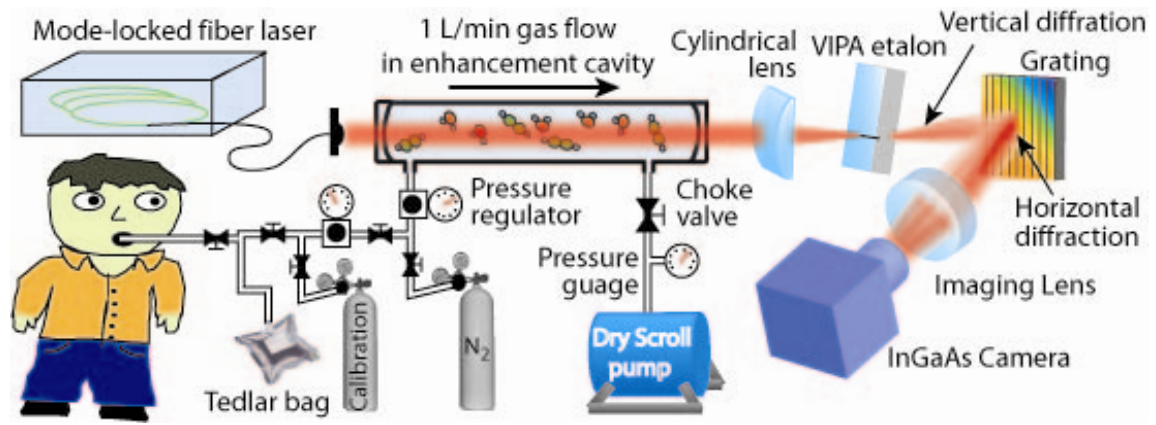

Fig. 1

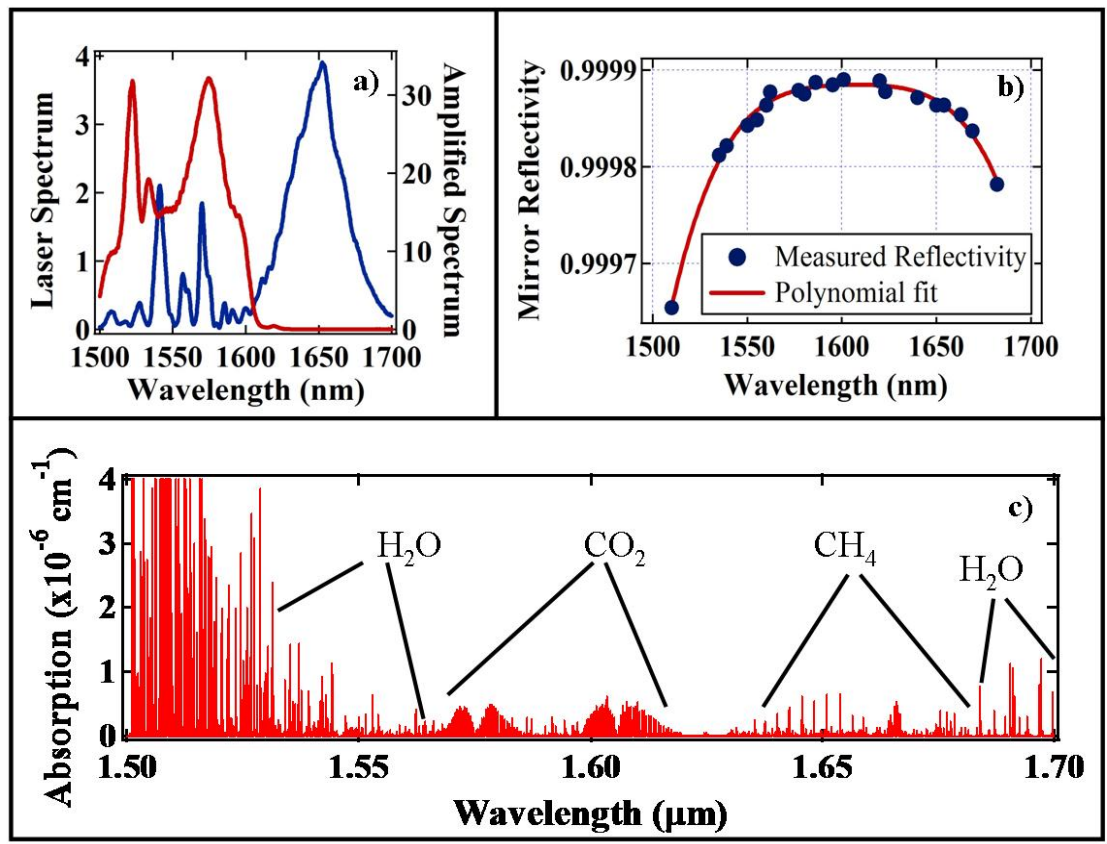

Fig. 2

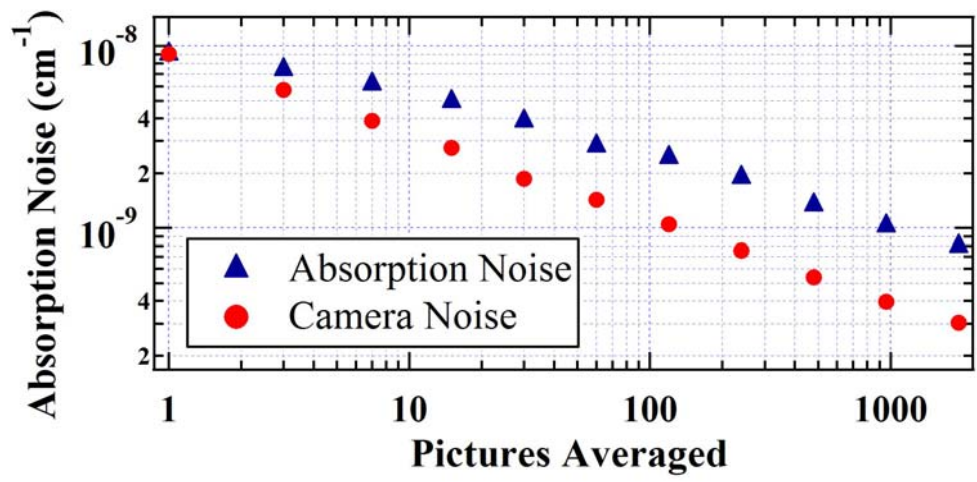

Fig. 3

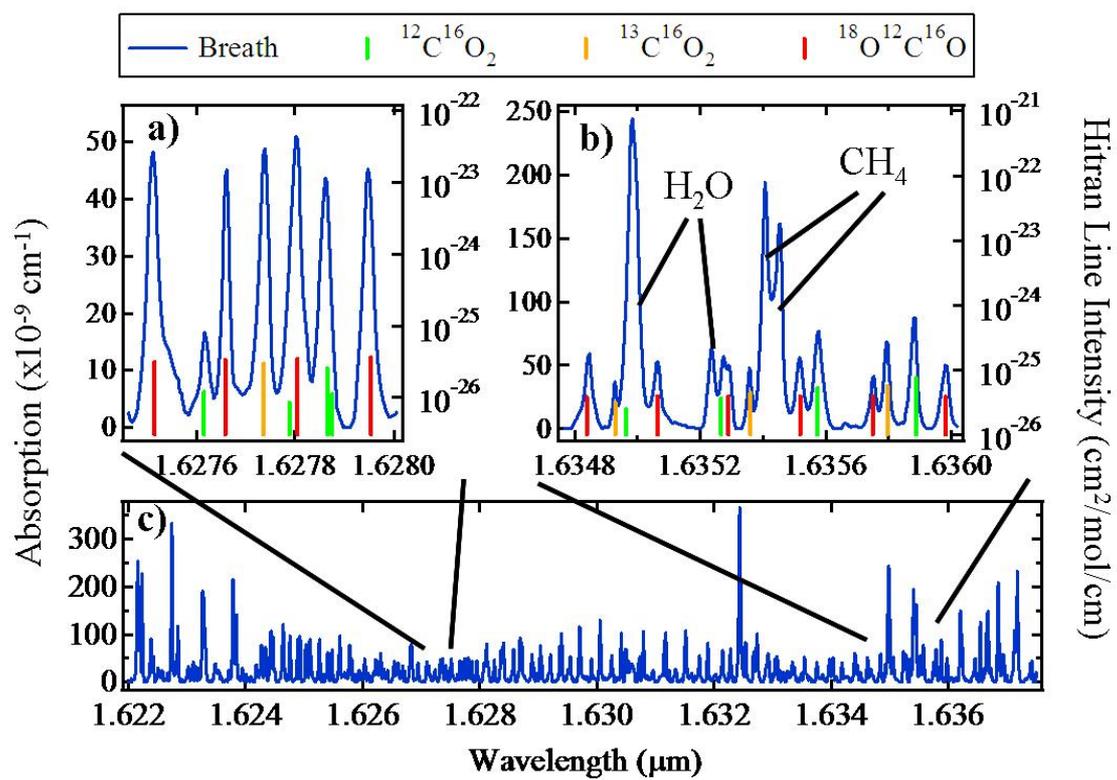

Fig. 4

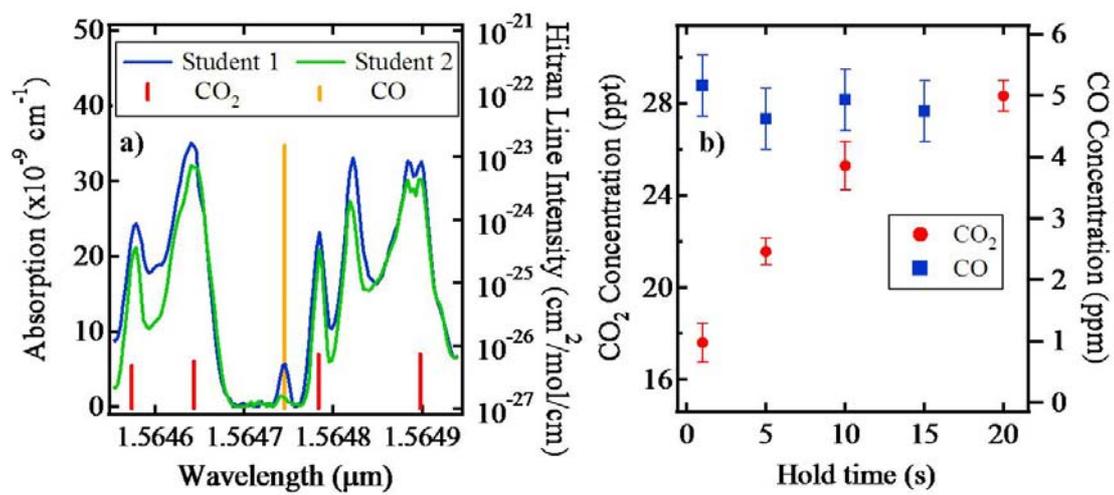

Fig. 5

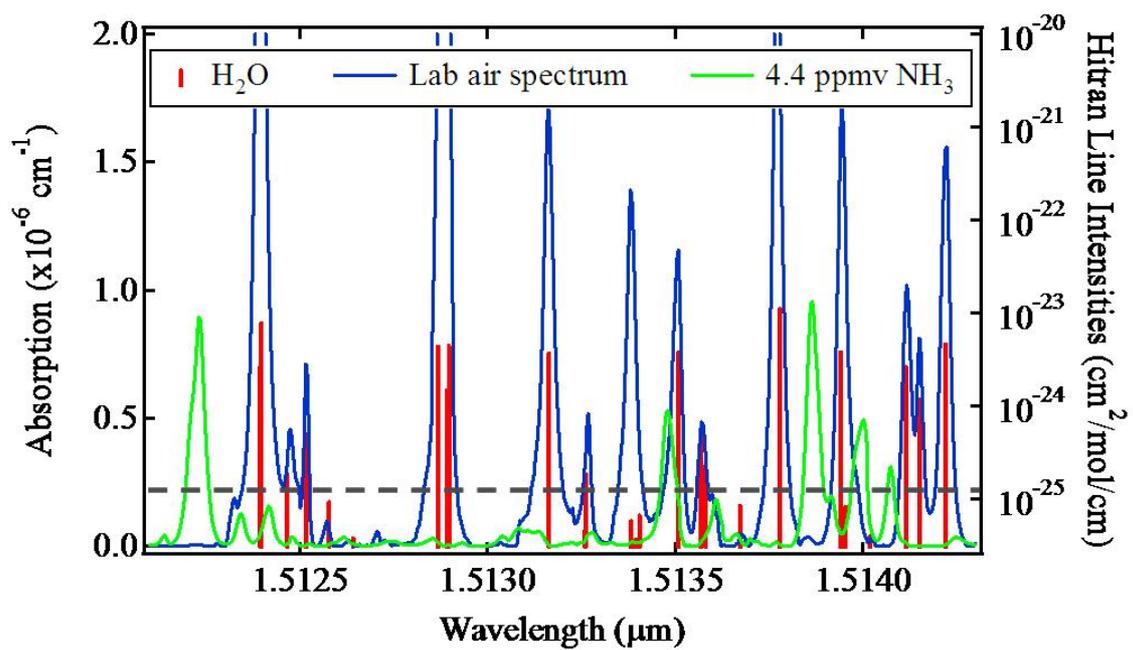

Fig. 6

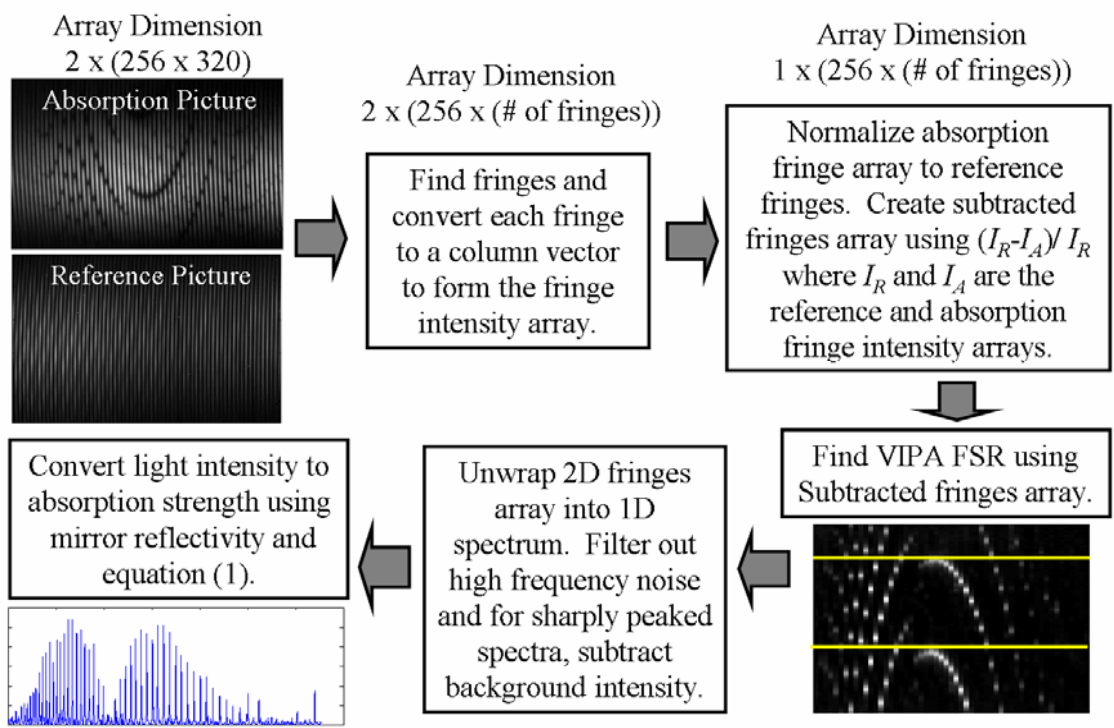

Fig. 7